# Proton-Transfer Ferroelectrics with Exceptional Switching Endurance

Bibek Tiwari[1], Yuanyuan Ni[1], and Xiaoshan Xu[1,2]
[1]Department of Physics and Astronomy, University of Nebraska-Lincoln, Nebraska 68588, USA.
[2]Department of Physics and Astronomy and the Nebraska Center for Materials and Nanoscience, University of Nebraska-Lincoln, Lincoln, Nebraska 68588, USA
E-mail: xiaoshan.xu@unl.edu

**ABSTRACT:**

Reliable organic ferroelectrics for memory applications require extreme endurance under repeated electrical switching. Here we demonstrate exceptional fatigue resistance in highly crystalline 2-methylbenzimidazole (MBI) films grown by low-temperature deposition followed by restrained crystallization (LDRC) in a simple Pt/MBI/Pt capacitor geometry. Switching kinetics analyzed using the Kolmogorov–Avrami–Ishibashi (KAI) model reveal characteristic millisecond switching times and quasi-one-dimensional domain growth associated with proton transfer along hydrogen-bond chains. Guided by these kinetics, we implemented a stringent fatigue protocol designed to maximize switching stress, involving bipolar switching at ~$2E_c$ with 5 ms pulses, well beyond the characteristic switching time, for continuous operation over ~2 weeks. The remanent polarization exhibits only a minor wake-up (+10% within the first $10^4$ cycles) and ultimately returns to approximately its initial value after $10^8$ cycles, with testing limited by experimental duration rather than device failure. This robust endurance is achieved in an unengineered structure and contrasts with polymer ferroelectrics such as P(VDF–TrFE), where comparable performance typically relies on interfacial engineering. The combination of LDRC-enabled high crystalline and localized proton-transfer switching, which introduces minimal structural perturbation during polarization reversal, enables this outstanding fatigue tolerance and highlights MBI as a simple, fluorine-free platform for durable organic ferroelectric devices.

## Introduction

Reliable organic ferroelectrics are essential for flexible nonvolatile memories, sensors, and actuators[1,2], where devices must sustain billions of switching events without significant polarization loss. Polymer ferroelectrics such as PVDF and P(VDF–TrFE) remain the most widely studied organic systems, offering sizable polarization and convenient solution processability[3]. However, their long-term reliability is limited by polarization fatigue, which is strongly influenced by interfacial charge injection, charge trapping, and field-driven transport processes[4]. In fluorinated polymers, electron-induced C–F bond breaking can produce HF that accumulates in the device stack, leading to interfacial degradation, electrode delamination, and reduced switchable polarization[3,5–7]. Although optimized electrodes, dielectric interlayers, and tailored driving schemes have improved endurance, fatigue in polymer ferroelectrics remains closely tied to device architecture and interfacial chemistry[8].

These considerations have motivated increasing interest in molecular ferroelectrics, where polarization arises from well-defined intermolecular interactions rather than polymer chain conformations[9]. Hydrogen-bonded molecular crystals, such as polycarboxylic acids and benzimidazole derivatives, are particularly promising, as their ferroelectricity originates from proton transfer along ordered hydrogen-bond networks, enabling polarization reversal through localized proton motion rather than large-scale structural rearrangements[9–14]. This localized proton-transfer mechanism can, in principle, reduce structural disruption per switching event and thereby improve fatigue endurance, especially when combined with high-crystallinity films that minimize defect-mediated pinning. Yet, systematic fatigue studies on molecular proton-transfer ferroelectrics and their microscopic switching kinetics remain scarce.

Here we investigate ferroelectric switching and long-cycle endurance in films of highly crystalline 2-methylbenzimidazole (MBI), a prototypical hydrogen-bonded molecular ferroelectric in which polarization reversal occurs via proton transfer along N–H···N bonds[15–17] (**Fig. 1a**), grown by a low-temperature deposition followed by restricted crystallization (LDRC) process on interdigital Pt electrodes prepatterned on glass. We show that LDRC produces spherulitic films with micrometer-scale fibers and near single-crystal-level polarization. The polarization switching measured on the Pt/MBI/Pt simple structure reveals Kolmogorov–Avrami–Ishibashi type switching[18] and with Avrami index $n \approx 1$ and a Merz's law[19]–type dependence on the applied electric field, consistent with nucleation-dominated switching mechanisms. Building on this kinetic understanding, we design a stringent high-field fatigue protocol and demonstrate that MBI maintains its remanent polarization closer to the initial value after $10^8$ cycles, corresponding to ≈2 weeks of continuous high-field operation, highlighting exceptional fatigue resistance compared with conventional P(VDF-TrFE)-based ferroelectrics.

## High crystallinity in LDRC films

The MBI films examined in this study were deposited by physical vapor deposition directly onto interdigitated electrode (IDE) on glass substrates. This was achieved using a low-temperature deposition followed by restrained crystallization (LDRC) method at 240 K, following the procedures reported in earlier studies.[16,20,21] The out-of-plane X-ray diffraction (**Fig. 1b**) confirms that the thin film (7 μm) adopts the bulk polar monoclinic phase, with reflections matching the simulated and powder reference pattern (CCDC-909439). Direct optical microscopy images in **Fig. S1a** reveal spherulite-like features with apparent nucleation centers. The spherulitic nature is unambiguously confirmed by cross-polarized optical microscopy as shown in **Fig. 1c** and in **Fig. S1b**, well-defined Maltese cross patterns are observed within individual spherulites, evidencing radially oriented, optically birefringent crystalline structures characteristic of spherulitic growth. Complementary atomic force microscopy in **Fig. 1d** further resolves μm-sized rod-like crystallites within the spherulite fibers, with ~500 nm diameters and >1 μm length.

Ferroelectric switching is probed using an interdigital electrode geometry, as schematically shown in **Fig. 1e**. As shown in **Fig. 1a**, MBI crystallizes in a bipolar ferroelectric structure composed of two molecular layers per unit cell; within the A and B layers the hydrogen-bond chain defines the polar axis along the $[10\bar{1}]$ and the $[101]$ directions respectively. This makes the polarization possible along any direction within the *a*-*c* plane. As shown in **Fig. 1f** (PE-loops for other frequencies and temperatures are shown in **Fig. S3**), well-defined P–E hysteresis loop with clear remanent polarization ($P_r$) and coercive fields is measured at 10 Hz. Notably, the $P_r$ value is comparable to the single-crystal values ($P_r \approx 5.2$ μC/cm$^2$ at 300 K, 0.2 Hz)[9,11,14,16], reflecting the large crystallite size and bipolar nature of MBI.

Taken together, these results indicate that the LDRC process produces highly crystalline MBI films. The combination of well-defined spherulitic morphology, large crystalline fibers, and near single-crystal-level ferroelectric polarization provides an excellent platform for investigating polarization switching dynamics and fatigue behavior.

## Switching dynamics and KAI behavior

To determine the switchable polarization, measurements were performed using a ferroelectric Tester following the pulse sequence shown schematically in **Fig. 2a**. The first (positive) voltage pulse initializes the sample into a well-defined polarization state. The second (negative) pulse, characterized by a pulse width ($t_p$) and pulse voltage ($V_p$) is applied to induce reverse polarization switching. A delay time $\Delta t$=100 ms is maintained between successive pulses to minimize transient effects. The third (positive) pulse is used to measure the polarization switched by the second pulse. However, this signal ($P^*$) contains contributions from non-ferroelectric effects such as leakage current and capacitive charging. To account of these contributions, a fourth (positive) pulse is applied to record the non-switchable polarization ($P^\wedge$). The net switchable polarization is then obtained as $P_{sw} = P^* - P^\wedge$ for a given pulse voltage and pulse width.

**Fig. 2b** shows the switchable polarization ($P_{sw}$) as a function of pulse-width ($t_p$) on a logarithmic timescale for several applied electric fields measured at room temperature (~295 K) for various electric fields. Clearly, increasing the applied voltage (electric field) leads to faster switching. For high electric field and long pulse widths, $P_{sw}$ saturates at ~2.8 μC/cm$^2$. Below 112.5 kV cm$^{-1}$, no clear saturation of polarization is observed up to 1000 ms, suggesting longer switching time. This behavior agrees with typical ferroelectric switching models in which stronger electric fields reduce the characteristic switching time by enhancing domain nucleation and domain-wall motion[22–27].

The experimental data were analyzed by fitting to the Kolmogorov–Avrami–Ishibashi (KAI) model as given by **Eqn. (1)**,

$$\Delta P_{sw}(t) = 2P_s\left[1 - \exp\left(-\left(\frac{t}{t_0}\right)^n\right)\right] \cdots\cdots\cdots\cdots (1)$$

where $P_s$ is the saturation polarization, $n$ is the Avrami index and $t_0$ is the characteristic switching time. The baseline value of polarization was estimated based on measurements with lowest electric field and shortest pulse width (10 ms, limited by the instrument) with extrapolation to lower time scale.

The strong agreement between the experimental results and the KAI model shown in **Fig. 2b** indicates that the polarization switching in the MBI film is governed by nucleation-and-growth dynamics. KAI theory was originally formulated for displacive inorganic ferroelectrics to describe an "ideal" case of switching with a narrow distribution of nucleation time and domain wall velocity. Ideal KAI behavior is typically restricted to large single crystals with spatially homogeneous switching; it is less seen in thin films due to contributions from interfaces, grain boundaries and defects. Therefore, the excellent fit in **Fig. 3b** suggests high crystallinity inside the large (μm-sized) rod-shaped crystallites (**Fig. 1d**), indicating that the grain boundaries contributions to the switching kinetics is unimportant.

### 1D Switching and Merz-Type Kinetics

To understand the role of thermal activation in polarization switching, the dependence of the switchable polarization on pulse width was investigated over the temperature range of 250-295 K with similar electric field (see **Fig**.**S4** for detail $P_{sw}$ at several electric fields for each temperature). **Fig. 3** illustrates the temperature-dependent polarization switching dynamics of the sample. At both electric fields in **Figs. 3a** and **b**, an increase in temperature results in a systematic shift towards smaller switching time. This aligns with the thermally activated nature of domain nucleation and growth, where increased thermal energy yields a higher probability of overcoming the switching barrier, as observed in polymeric systems like PVDF[25].

The Avrami index $n$ in **Eqn. 1** serves as an indicator of the effective dimensionality of domain-growth kinetics and the nature of the nucleation process; the extracted values are plotted in **Fig. 4a**. Across all temperatures, $n$ remains close to 1 and increases modestly with electric field, suggesting that stronger fields promote a transition from predominantly one-dimensional to slightly more multidimensional switching pathways. Intrinsically, $n \approx 1$ reflects the anisotropy of proton-transfer switching in MBI. Like shown in **Fig.1a**, proton displacement occurs preferentially along H-bonded N-N…N chains, creating quasi 1D switching pathways at low fields. Lateral expansion of switchable regions requires correlated molecular reorientation across neighboring chains which is energetically less favorable thereby leading to domain growth effectively constrained to 1D at lower fields. The near-unity values confirm that polarization switching tracks the hydrogen-bond chains, another indication of high crystallinity, since abundance of defects may undermine the 1D behavior. At higher fields, additional nucleation sites become active for transverse nucleation, enabling a more extended growth geometry reflected in the rising $n$ values. Conversely, $n$ decreases systematically with decreasing temperature, indicating a suppression of nucleation activity and increasingly constrained domain-wall motion as thermal energy is reduced.

The characteristic switching time $t_0$, extracted from the KAI model fit, is plotted in **Fig. 4b** as a function of 1/E for all temperatures to examine its compliance with Merz's empirical law,

$$t_0 = t_\infty \exp\left(\frac{\alpha}{E}\right) \cdots\cdots\cdots\cdots\cdots\cdots\cdots\cdots\cdots (2)$$

where α is an activation field associated with the energy barrier for domain reversal and $t_\infty$ is the switching time at an infinite field. The linear behavior observed in the semi-log plot confirms that the switching dynamics follow Merz-type kinetics across the entire temperature range.

The nearly parallel lines for different temperatures indicate that the activation field $\alpha$ is only weakly dependent on temperature within this range. Quantitively, from the fits shown in **Fig. 4b,** the extracted values of $t_\infty$ and α at 295 K are approximately 0.14 ms and 0.08 GV $m^{-1}$ respectively. This slow switching time scale is also evident in BTA-C18 type systems[27] but the slow nature comes here with long carbon chains with higher masses favoring delayed switching unlike for BTA-C6 (~μs). In MBI, the strong correlation between the proton position and the π-electron states in the C-N rings could slow down the proton transfer in the collective macroscopic behavior[10]. In addition, this $\alpha$ could also be overestimated due to the interdigitated electrode nature of in-plane electric field measurement for order less than that observed in PVDF type polymeric systems[24,28]. However, in contrast to PVDF systems, where switching involves conformational changes of polymer chains[29], MBI switching is driven by localized proton transfer within rigid molecular frameworks. This distinctive nature also might favor the lower activation fields in general.

Upon lowering the temperature (**Fig. S5**), $\alpha$ shows a slight increase, while $t_\infty$ exhibits a non-monotonic dependence. The dominant effect of reduced temperature is an upward shift of the curves, indicating an overall increase in $t_0$ due to suppressed thermal activation. Consistently, the weak temperature dependence of the activation field is evident from the $\alpha$ versus $1/T$ plot **(Fig. S5a)**, whereas $t_\infty$ retains its non-monotonic behavior (**Fig. S5b**). Thus, in proton-transfer ferroelectrics, the activation field extracted from Merz law reflects the field required to distort/tilt proton double-well within H-bond. Unlike displacive ferroelectrics, where barrier is dominated by collective ionic motion, the barrier in MBI arises from localized proton transfer, the fundamental switching unit is a proton hopping between adjacent sites[10]. This localized nature of the switching unit is possibly weakly coupled to the lattice, leads to a modest temperature dependence of the activation field, so the weaker temperature dependence of the activation field is plausible. But the nucleation sites are inherently temperature dependent as also discussed in TA-NLS[30] model, causing the strong temperautre dependence of the characteristic switching time.

### Exceptional Fatigue resistance

Leveraging our understanding of the switching kinetics, we designed a stringent testing protocol to evaluate the ultimate fatigue resistance of the MBI films. First, we utilized a high bipolar voltage pulse of 180 V—approaching the maximum limit of the tester—which corresponds to a substantial applied electric field of approximately 450 kV/cm (approximately $2 \times E_c$). Based on the $t_\infty$ and α values extracted from the Merz law fits (**Fig. 4a**), the characteristic switching time $t_0$ at this specific field is calculated to be 0.82 ms. Consequently, we selected a pulse width of 5 ms for the fatigue cycling. This pulse duration ($>>t_0$) ensures complete polarization reversal during every cycle, avoiding the partial-switching artifacts which may inflate cycle counts in high-frequency endurance testing.

The fatigue measurements were performed on a 2 μm sample continuously for approximately two weeks, reaching cumulative switching cycles up to $10^8$ under these high-field conditions. To accurately account for non-ferroelectric contributions such as leakage current, the switchable polarization was measured intermittently using the positive-up-negative-down (PUND) method.

As shown in **Fig. 5a**, the remanent polarization ($P_r$) initially increases and saturates after $10^3$ cycles. This is indicative of a "wake-up" effect, a well-known phenomenon in highly crystalline ferroelectrics driven by the field-induced de-pinning of domains, relaxation of internal stresses from the LDRC growth process, and the stabilization of interfacial charge screening. Following this initial wake-up phase, the polarization remains remarkably constant up to $\sim 10^7$ cycles, demonstrating excellent fatigue endurance.

At higher cycle numbers, a slight degradation emerges; however, the overall reduction in $P_r$ from its peak value remains below 10%. Ultimately, at $10^8$ cycles, the polarization stabilizes at a value nearly identical to the pristine, uncycled state. The corresponding $P$–$E$ hysteresis loops in **Fig. 5b** corroborate this trend, showing the initial increase followed by the gradual decrease of $P_r$, accompanied by a consistent, modest increase in the coercive field ($E_c$) with extended cycling, which suggests progressive domain wall pinning.

These results unequivocally demonstrate the robust fatigue tolerance of the proton-transfer MBI ferroelectric films. Unlike fluorinated polymers such as P(VDF-TrFE)[6,7], where extended high-field cycling frequently leads to irreversible chemical degradation (e.g., C–F bond cleavage) and severe macroscopic fatigue, the localized proton-transfer mechanism within the rigid MBI molecular framework intrinsically resists structural damage. Maintaining stable, fully saturated switching over $10^8$ cycles during two weeks of continuous ambient operation establishes LDRC-grown hydrogen-bonded molecular crystals as a highly durable alternative for long-lasting organic electronics.

**Conclusions:**

In conclusion, we have established a structure-property framework for achieving robust fatigue resistance in hydrogen-bonded molecular ferroelectrics. Using the low-temperature deposition followed by restrained crystallization (LDRC) method, we fabricated highly crystalline 2-methylbenzimidazole (MBI) films exhibiting large spherulitic domains and near single-crystal-level remanent polarization. The polarization switching follows near-ideal Kolmogorov–Avrami–Ishibashi (KAI) kinetics with an Avrami index $n \approx 1$, consistent with quasi-one-dimensional switching along the hydrogen-bond chains. Guided by the extracted switching kinetics, we implemented a stringent fatigue protocol designed to maximize switching stress, employing electric fields up to 450 kV/cm and pulse widths well beyond the characteristic switching time. Under these conditions, a simple Pt/MBI/Pt capacitor structure was continuously cycled for more than $10^8$ switching events over approximately two weeks. Remarkably, the remanent polarization after $10^8$ cycles remains comparable to its initial value, indicating negligible polarization fatigue. The exceptional endurance likely arises from the combination of high crystallinity enabled by LDRC growth and the localized proton-transfer switching mechanism in MBI, which introduces minimal structural distortion during polarization reversal. These results demonstrate that molecular ferroelectrics can achieve outstanding switching durability even in simple metal–ferroelectric–metal architectures, highlighting hydrogen-bonded systems such as MBI as promising platforms for reliable, low cost, and sustainable organic ferroelectric devices.

**Methods:**

The MBI films were deposited by thermal evaporation using a physical vapor deposition system (EvoVac, Angstrom Engineering) operated under high vacuum (~$10^{-6}$ Torr). A schematic of the modified deposition setup is provided in our earlier work (**Fig. 1b**[20]). Film growth was carried out on glass substrates patterned with Pt interdigitated electrodes (IDEs) (G-IDEPT5, Metrohm, USA) featuring a 10 µm period and 4 µm electrode spacing (**Fig. S9a**[20]). The IDEs were interfaced with a Radiant Precision RT66C ferroelectric tester to perform electrical polarization measurements and switching dynamics analysis. Depositions were conducted at a substrate temperature of 240 K, regulated by a Lakeshore temperature controller employing liquid-nitrogen circulation to achieve cryogenic conditions. Film thickness was monitored in situ using an integrated quartz crystal microbalance (QCM) and subsequently calibrated using several test samples, revealing a deviation of approximately 100 nm from the nominal thickness. The spherulitic morphology was examined using a reflective cross-polarized optical microscope (Nikon Eclipse L200N). Microscale surface topography was characterized with a Keyence laser scanning microscope, while nanoscale features were analyzed via atomic force microscopy (Bruker Icon) operated in ScanAssyst Peak Force Tapping mode. Structural characterization was performed using a Bruker D8 diffractometer equipped with a 2D detector.

**Data availability**

Data available on request from the authors. Conflicts of interest: There are no conflicts to declare.

**Acknowledgements**

This research was primarily supported by the UNL Grand Challenges catalyst award entitled Quantum Approaches addressing Global Threats. This work was also supported in part by the Nebraska Center for Energy Sciences Research (NCESR). The research was performed in part at the Nebraska Nanoscale Facility: National Nanotechnology Coordinated Infrastructure and the Nebraska Center for Materials and Nanoscience, which are supported by the NSF under Grant No. ECCS-2025298 and the Nebraska Research Initiative.

## Main Text Figures

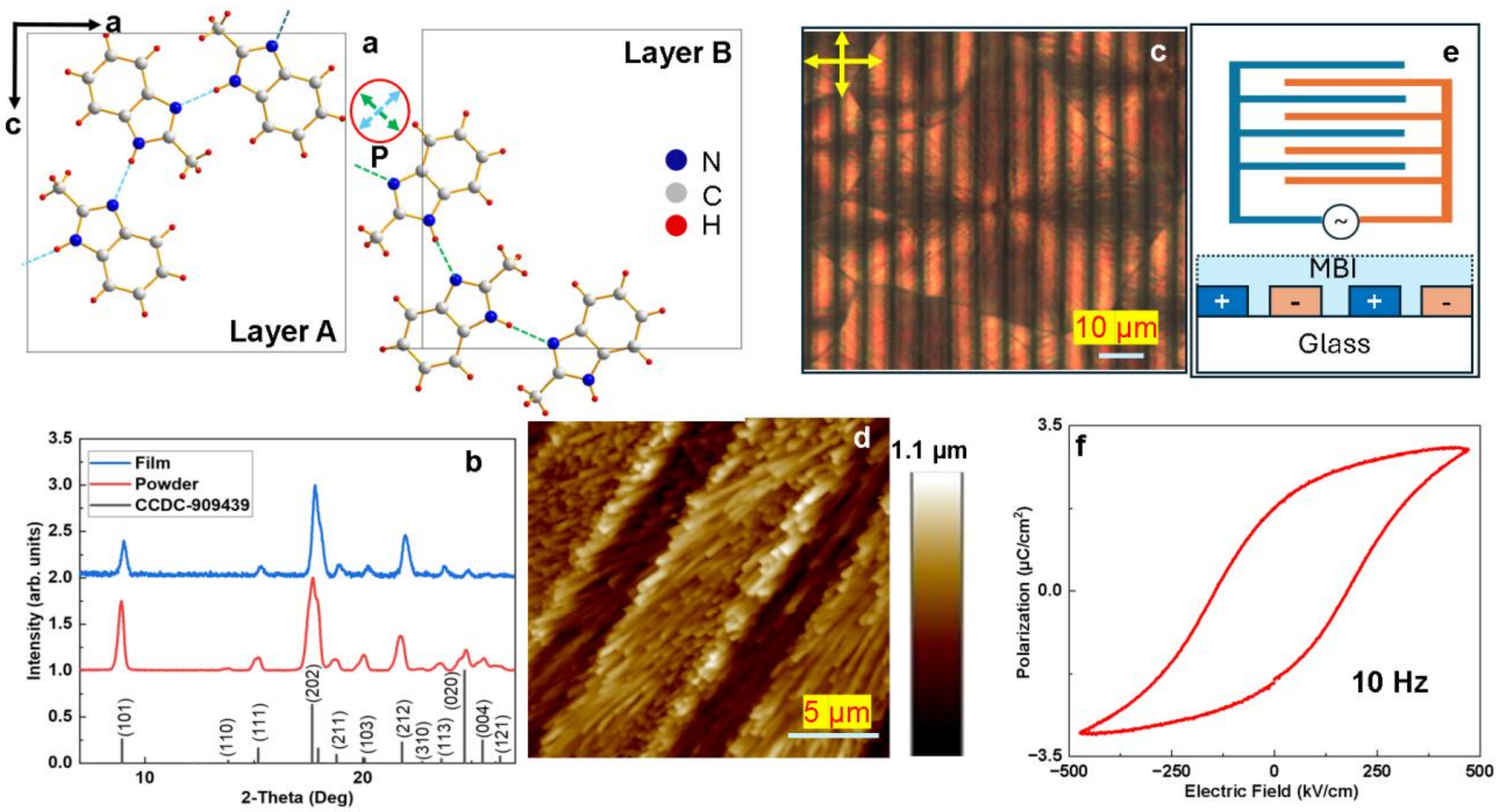

**Figure 1. Structural origin, microstructure, and ferroelectric response of the molecular MBI organic film.** (a) Polar crystal structure highlighting hydrogen-bonded molecular layers and proton-transfer-induced spontaneous polarization (shown by green and blue arrows and dots, polarization direction shown inside red circle). (b) X-ray diffraction patterns of the thin film compared with powder and reference data, confirming phase purity and preferred orientation. (c) Cross-polarized optical microscopy image showing anisotropic ferroelectric domain morphology and Maltase-cross patterns. The stripe patterns are from the interdigital electrodes (see panel e). (d) Atomic force microscopy topography revealing fiber-like crystallites. (e) Schematic of the interdigitated electrode (IDE) with gap size of 4 $\mu$m and period of 10 $\mu$m configuration used for in-plane electrical measurements. (f) Polarization–electric field (P–E) hysteresis loop measured under IDE geometry at 10 Hz, demonstrating switchable ferroelectric polarization driven by proton transfer within hydrogen bonds. Film thickness is 7 µm.

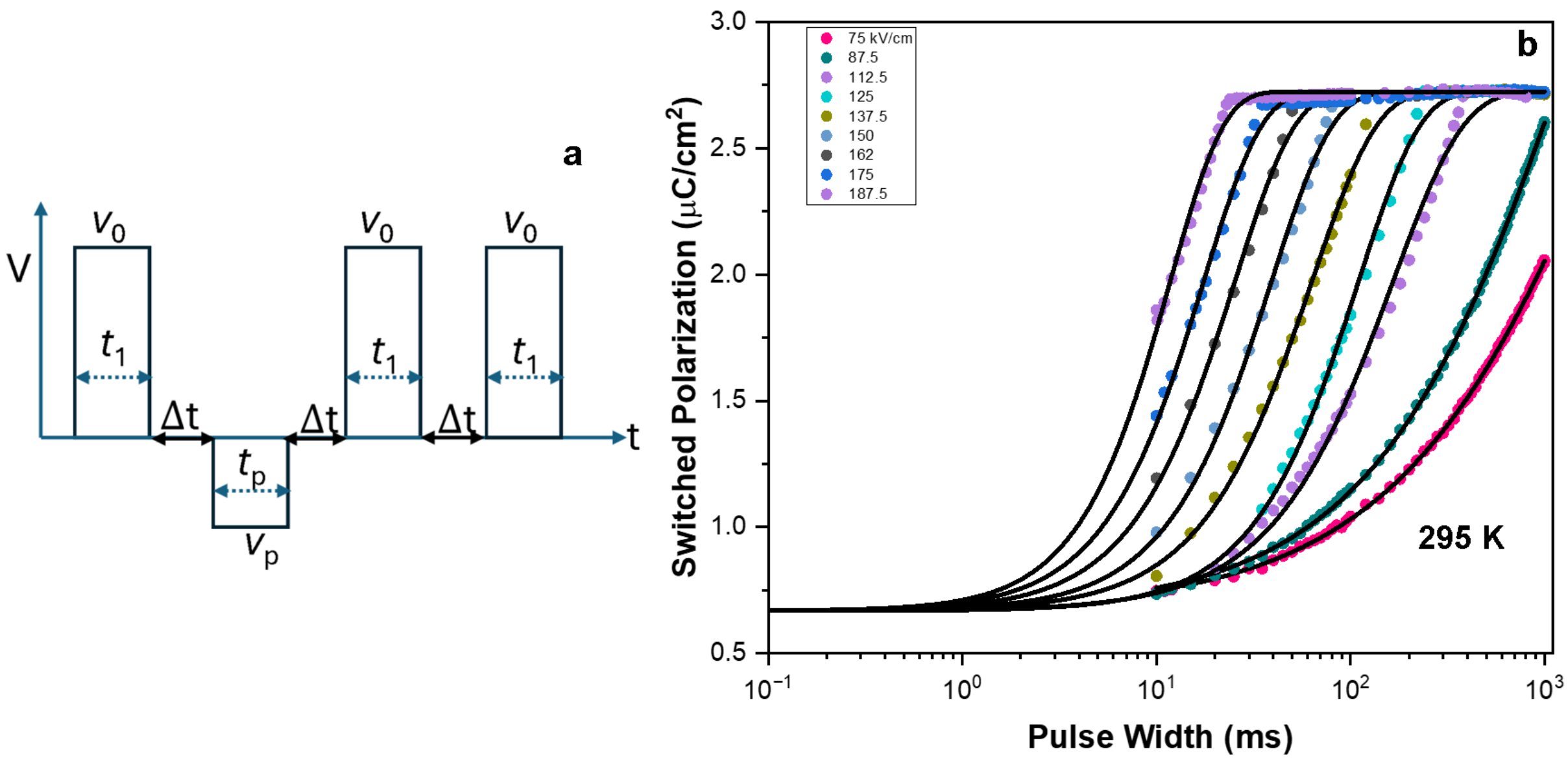

**Figure 2**. (a) The pulse sequence scheme for switching dynamics measurement where $V_0$ =190 V, $\Delta t$=100 ms, and $t_1$=1000 ms, $V_p$ and $t_p$ are varied. (b) Switchable polarization ($P_{sw}$) vs. Pulse width ($t_p$) at 295 K for various electric fields $E$ (by varying $V_p$) and solid lines are fit to KAI model. Film thickness is 7 μm.

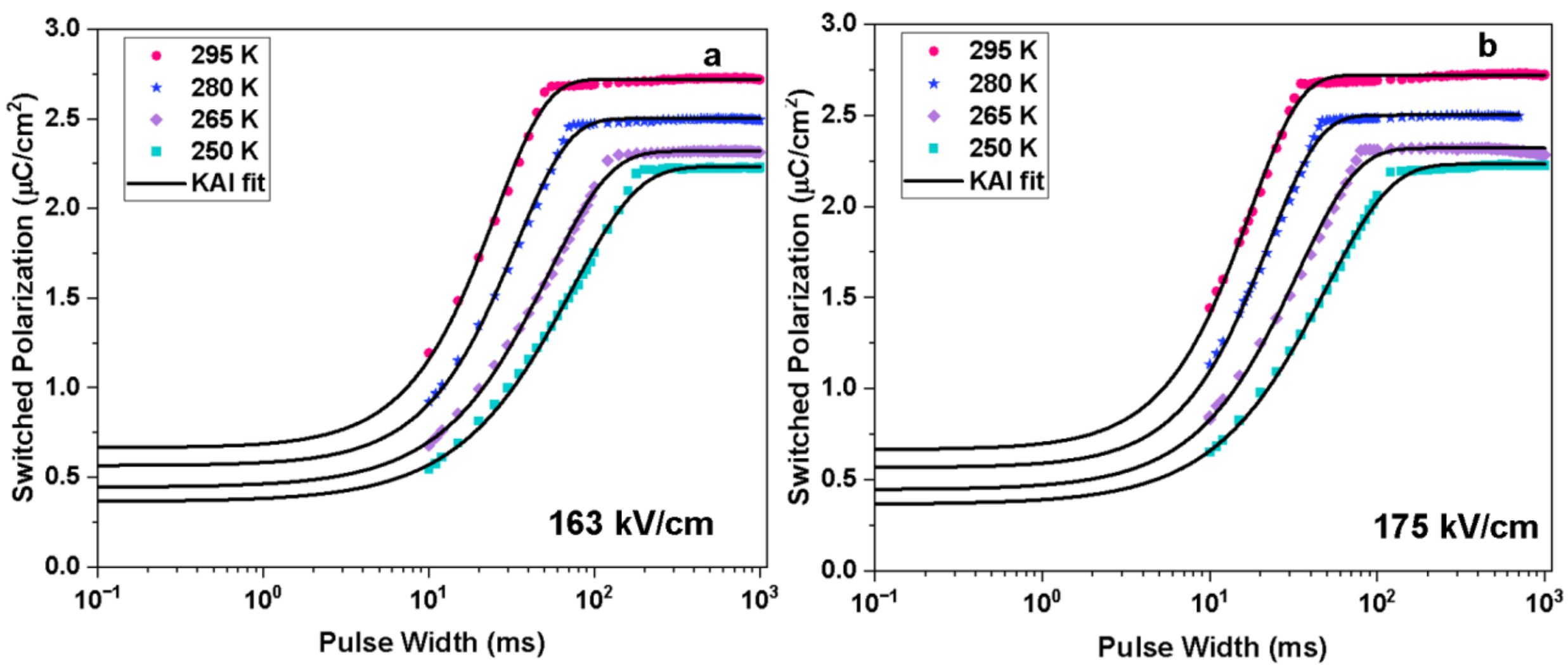

**Figure 3**. (a) Temperature dependence of switchable polarization at a fixed electric field of 163 kV/cm (Several field dependences at a given temperature is shown in S4), illustrating slower switching at lower temperatures and solid line is a fit to KAI model. (b) Similar temperature dependence graph at a fixed electric field of 175 kV/cm. Film thickness is 7 μm.

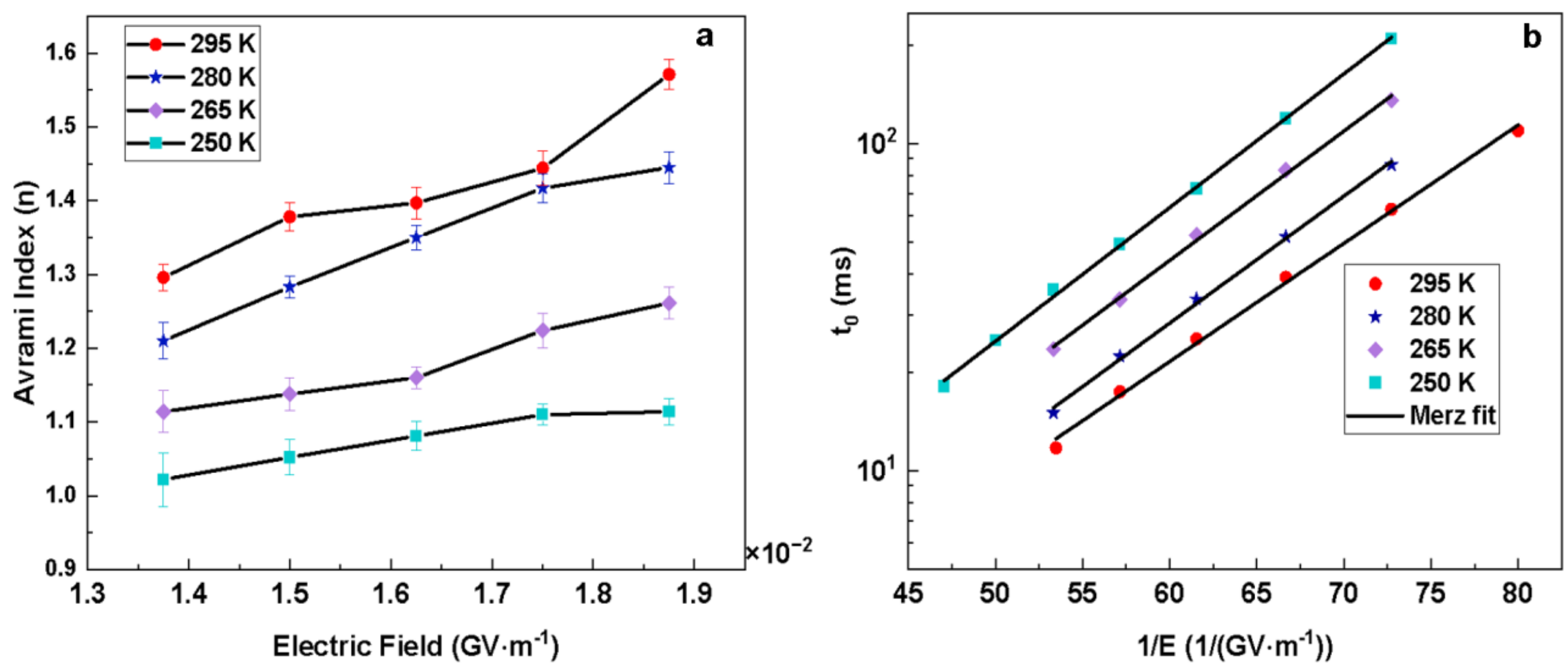

**Figure 4. KAI model parameters**. (a) Avrami index $n$ as a function of electric field at the same temperatures range, indicating field and temperature-dependent domain-growth kinetics. (b)Plots of characteristic switching time $t_0$ $t_0$versus inverse electric field $1/E$ for 250–295 K, showing linear behavior consistent with Merz's law (solid lines are fits). Film thickness is 7 μm.

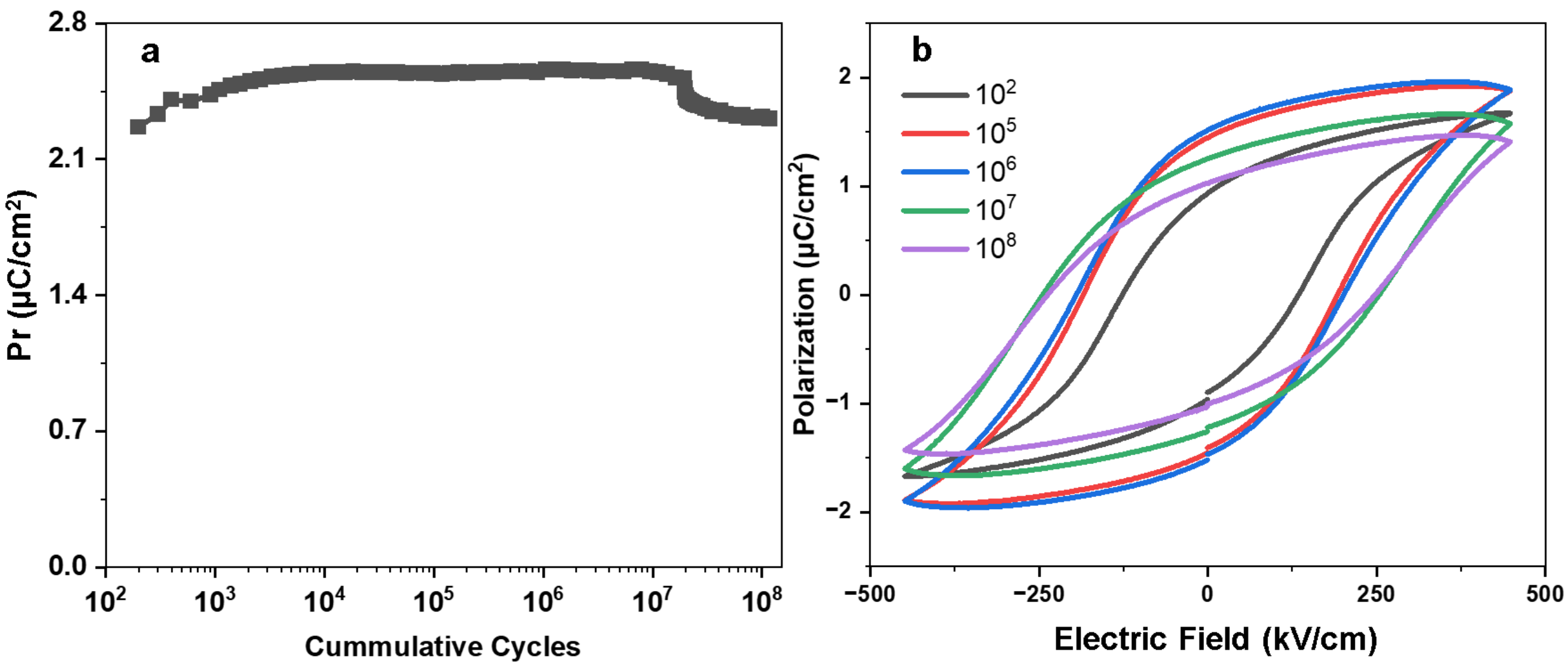

**Figure 5**. **Fatigue behavior of the MBI ferroelectric film measured using bipolar pulses of 180 V amplitude and 5 ms width.** (a) Remanent polarization $P_r$ as a function of cumulative switching cycles. (b) Corresponding P–E hysteresis loops at selected cycle numbers, showing the evolution of polarization switching during repeated high-field cycling. Film thickness is 2 μm.

**Supplementary Figures:**

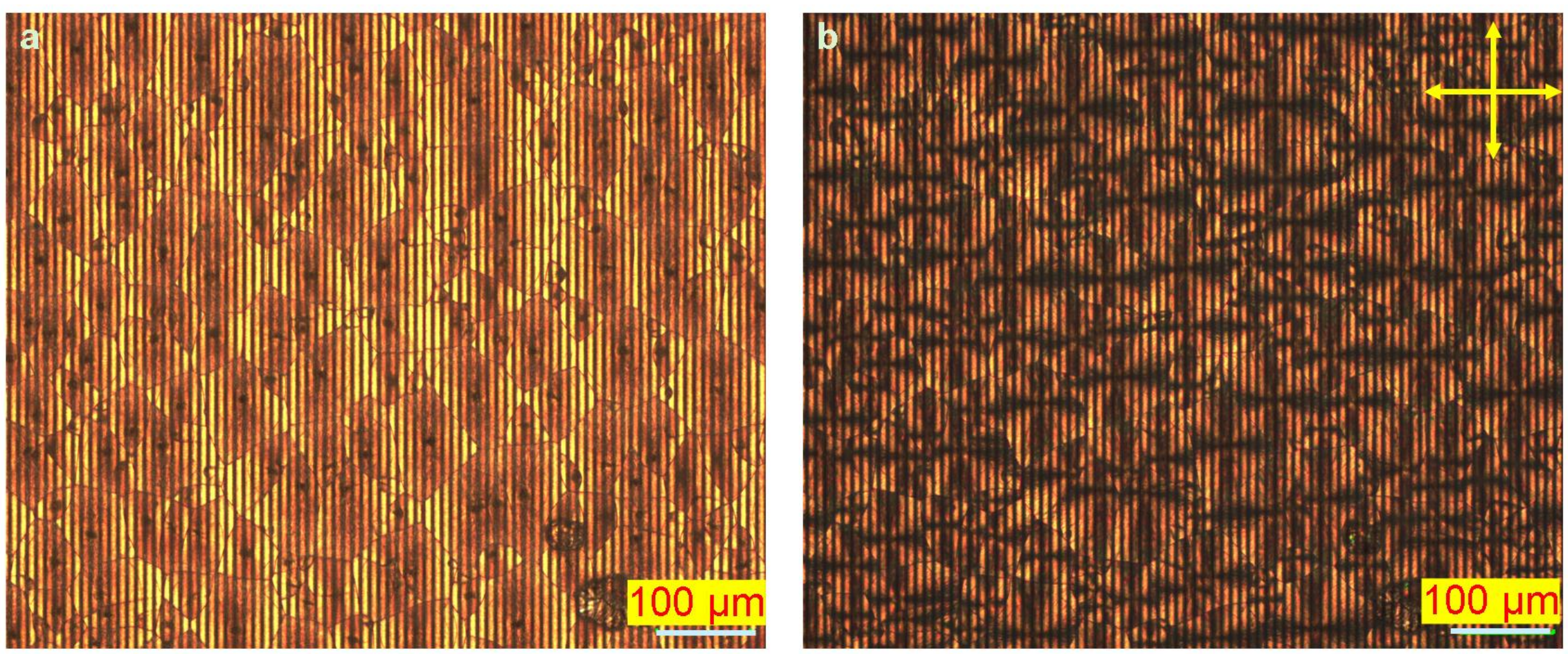

**Figure S1.** Optical Images of MBI grown on an IDE substrate. (a) and (b) shows the images in $0^0$ and $90^0$ (crossed position) between the polarizer and analyzer. The black spots on (a) are typical centers of the spherulitic morphology.

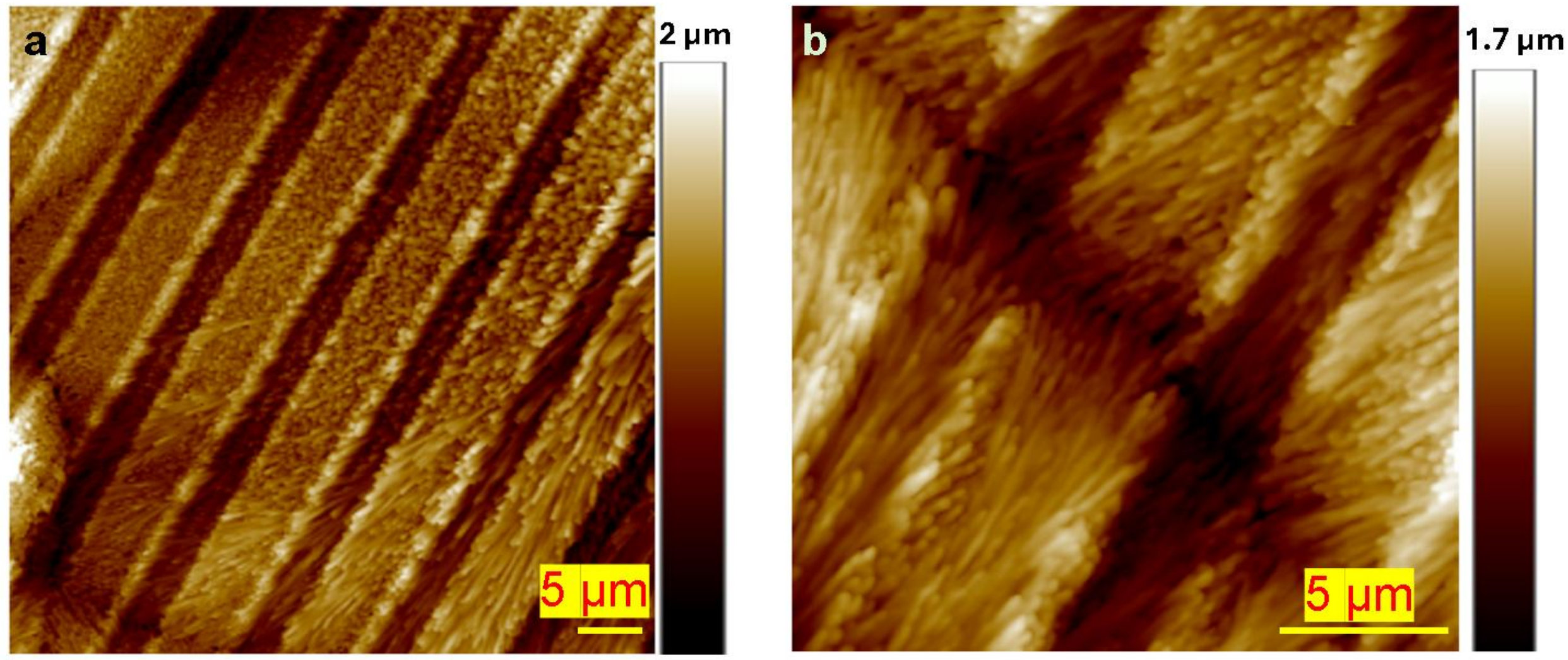

**Figure S2.** AFM images at different scales (a) 50 μm x 50 μm (b) 5 μm x 5 μm. RMS roughness is ~80 nm within one finger.

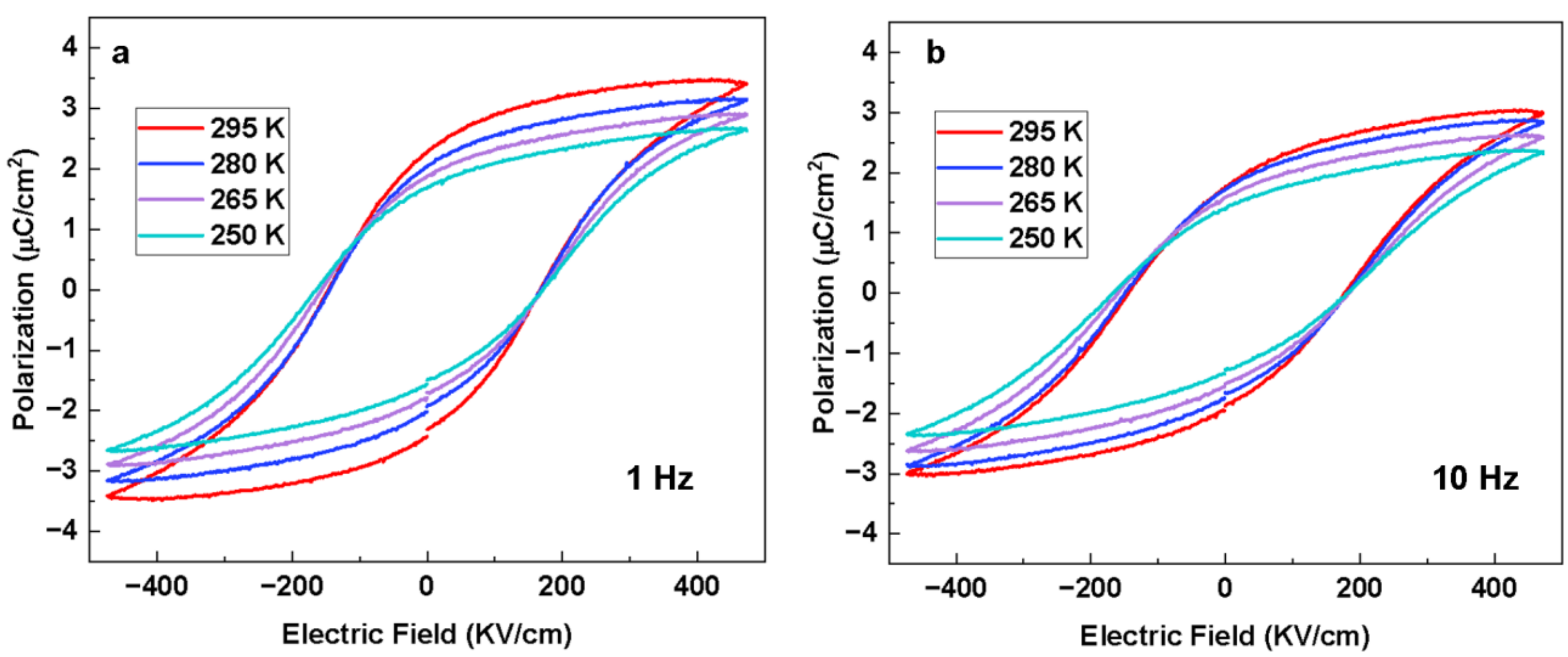

**Figure S3.** Polarization hysteresis loops of MBI at different temperatures at 1Hz (a) and 10 Hz (b)

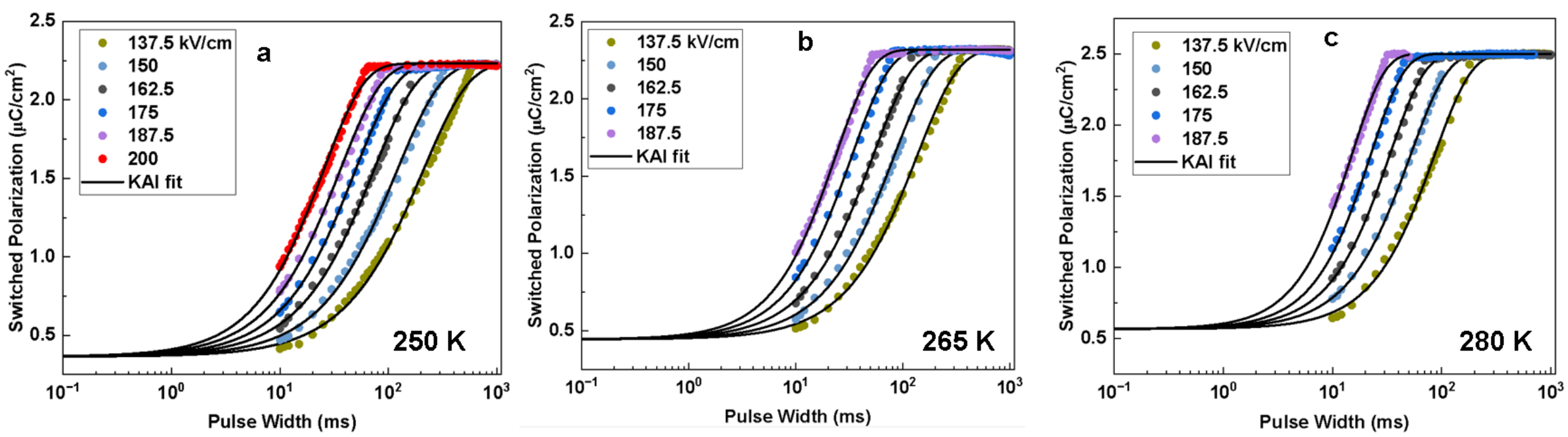

**Figure S4**. (a) Plot of switched polarization at several electric fields at temperatures: 250 K (a), 265 K (b) and 280 K (c) while the solid line is a fit to KAI model.

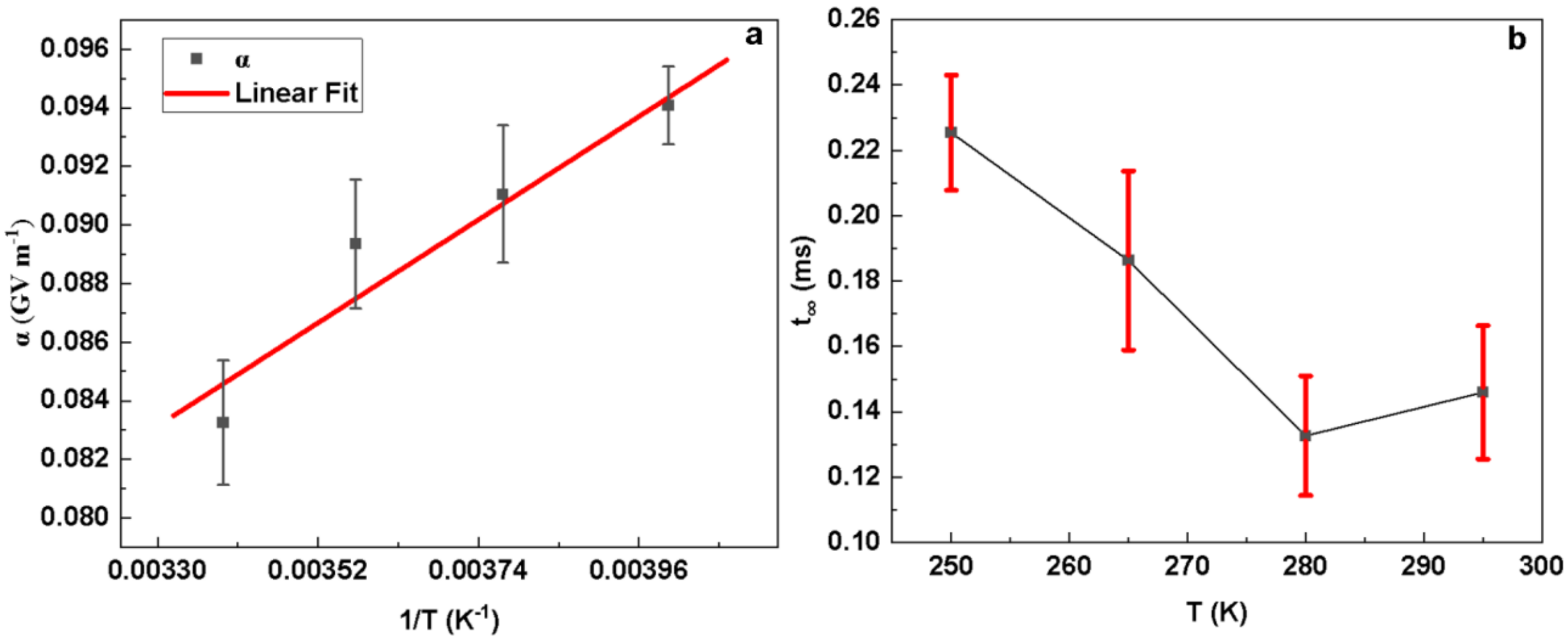

**Figure S5**. (a) Temperature dependence of the Merz law parameter α, plotted as a function of inverse temperature (1/T) with a linear fit. (5b) shows the corresponding limiting switching time at infinite fields ($t_\infty$) as a function of temperature with dashed line being a guide to the eye.